\documentclass[11pt,a4paper]{article}
\usepackage{jcappub}

\usepackage{color}
\usepackage[normalem]{ulem} 

\title{Light propagation in the averaged universe}

\author{Samae Bagheri,}
\author{Dominik J Schwarz}
\affiliation{Fakult\"at f\"ur Physik, Universit\"at Bielefeld,\\ Universit\"atstr. 25, Bielefeld, Germany}

\emailAdd{s\_bagheri@physik.uni-bielefeld.de}
\emailAdd{dschwarz@physik.uni-bielefeld.de}

\abstract{
Cosmic structures determine how light propagates through the Universe
           and consequently must be taken into account in the interpretation of observations.
           In the standard cosmological model at the largest scales, such structures are either ignored
           or treated as small perturbations to an isotropic and homogeneous Universe. This isotropic and 
           homogeneous model is commonly assumed to emerge from some averaging process at the 
           largest scales. We assume that there exists an averaging procedure that preserves the causal 
           structure of space-time. Based on that assumption, we study the effects of averaging the 
           geometry of space-time and 
           derive an averaged version of the null geodesic equation of motion.
           For the averaged geometry we then assume a flat Friedmann-Lema\^{i}tre (FL)
           model and find that light propagation in this averaged FL model is not given by null geodesics
           of that model, but rather by a modified light propagation equation that contains an  
           effective Hubble expansion rate, which differs from the Hubble rate of 
           the averaged space-time.}
          
\begin{document}

\maketitle

\section{Introduction}

The standard model of modern cosmology describes large-scale structures as 
perturbations of an isotropic and homogeneous Universe, the 
Friedmann-Lema\^{i}tre (FL) model. Measurements of the cosmic microwave 
background (CMB) and many other observations confirm that the standard model is a 
good description of the Universe. Despite of its success, the FL model is only a large-scale 
approximation to highly nonlinear structures at small scales. Consequently, one can ask 
how to justify this high degree of symmetry at the largest scales and how to connect the 
smallest scales to the largest ones. Eventually, we must not ignore the effects of local 
inhomogeneities from which an averaged space-time with certain symmetries seems to emerge.  
By \textit{local} we refer to scales on which gravitationally bound structures exist, 
i.e.~from $\sim 100$ Mpc down to the Planck scale. Above the $100$ Mpc, the 
Universe appears to be statistically homogeneous and isotropic, but on smaller scales, 
unlike the FL model, it is inhomogeneous. For testing large-scale homogeneity, several 
tests have been applied to the data from the Sloan Digital Sky Survey \cite{Hogg} and the 
WiggleZ Dark Energy Survey \cite{Blake}. In both cases a transition to homogeneity at 
scales of about $100$ Mpc is found. In this work, we are specifically interested on the effects of 
the local inhomogeneities at and below the $100$ Mpc scale on the propagation of light.
 
The averaging problem was introduced in general relativity by Shirkov and Fisher in 
1963 \cite{Fisher}. They proposed a space-time averaging procedure, but it was not 
covariant such that a tensor did not remain to be a tensor after applying averaging.
The issue was not very well known until 1984 when Ellis gave a 
description of the concept of \textit{backreaction} from small to large structures \cite{Ellis}.
The question was further considered by Futamase \cite{65,66} who
studied the gravitational correlation by employing the metric perturbations
and by Zotov and Stoeger \cite{129}, whose procedure was equivalent to the one
by Shirkov and Fisher, hence not covariant.

Two breakthroughs in the study of the averaging problem and backreaction were achieved 
by Zalaletdinov \cite{1992,1993} in a covariant and exact way and by Buchert 
\cite{19,buchert}, who restricted the problem to scalar quantities only.
A connection to dark energy has been proposed 
in \cite{Schwarz 2002}, \cite{rasanen}, \cite{Wiltshire111}, \cite{Wiltshire222}, and \cite{kolb}, 
who attempted to explain dark energy by means of a backreaction of small scale structures on the large 
scale evolution of the Universe, while many others like \cite{wald111} and \cite{wald222} were completely 
against that idea.
The question is still open. So far nobody could present a proof that would exclude this idea and 
nobody could prove that the backreaction effects are large enough to explain dark energy.
However, it seems to be generally accepted that backreaction effects cannot be neglected 
if one is interested in precision cosmology. The idea has been later on discussed in 
\cite{baz Coley, Kolb2010, R2006, schwarz1, schwarz2, Clarkson2} and many others.
 
The aim of this paper is to investigate the effects of averaging on light propagation 
in the Universe, or how to derive the equation of motion of light in an averaged 
description of the Universe from the null geodesic equation in the inhomogeneous universe.  
Therefore we ask if some effects can be seen in observations in the lumpy universe. 
For example how do averaged inhomogeneities affect the redshift of photons.
  
The motion of photons in an averaged geometry has already 
been studied in \cite{coley2} and in a more precise way in 
\cite{R2008, R2009}. Yet in a different approach using a gauge invariant formalism,
the averaged geometry on the past null cone
has been introduced \cite{veneziano1, veneziano2}.
This allows to average the luminosity-redshift relation 
\cite{veneziano3, veneziano4, veneziano5}. The study of light propagation in inhomogeneous
Swiss cheese models by simulating Hubble diagrams has been probed recently in \cite{Fleury1, Fleury2, Fleury3}.

Here we follow a new approach. We make the plausible assumption that an averaging procedure 
that respects the causal structure of space-time exists. Based on this and a second assumption specified 
in Sec.~3, we derive an effective equation for light propagation in an averaged Universe.

The central finding of our work is that photons in an averaged Universe follow a FL geodesic equation 
of motion, but with the Hubble rate replaced by an effective Hubble rate that does not coincide 
with the Hubble rate that one would infer from the averaging of the space-time itself. In contrast 
to many previous studies, this result is not based on a perturbative approach and does not make use of 
a toy model.

The work is organized as follows. 
In the next section we introduce the concept of a covariant averaging of a tensor and briefly 
discuss what has been achieved in the works by Zalaletdinov and Buchert. It is essential for 
our work that a covariant procedure to average a space-time metric exists. As we show below, 
it is irrelevant for our purpose how this is defined in detail. In the third section we derive our 
central result --- an effective equation for the propagation of light in an averaged space-time 
and in section four we evaluate that equation for a Universe that can be described by a FL model 
after averaging. The last two sections contain a discussion and a conclusion. 
  
\section{Averaging procedures of space-time}
 
Averaging involves the integration of tensors over a (space-time or spatial or null) volume $V$, 
and is not easily well defined, because the result can change by changing the coordinates and 
is typically not unique. For treating this problem one can define the covariant averaging of tensors 
via bilocal operators, as proposed by Zalaletdinov \cite{Zalal 2008,1992,1993}, or one can 
simplify the problem and consider only scalar quantities and average them, as has been first proposed 
by Buchert \cite{20,19,buchert}.

Besides the question of how to average a tensor, general relativity is non-linear in the components of the 
metric tensor $g_{\mu\nu}$, and thus in general for the Einstein tensor $G_{\alpha \beta}$
\begin{equation}
\langle G_{\alpha\beta} (g_{\mu\nu})\rangle \neq G_{\alpha\beta} (\langle g_{\mu\nu}\rangle), 
\end{equation}
where the brackets denote some averaging procedure.

In the procedure provided by Zalaletdinov one considers the bilocal extension of 
a tensor, e.g.~for a vector $P^\alpha(x)$ one defines its extension as 
$\tilde{P}^\alpha(x',x)=\mathcal{A}^{\alpha}_{\alpha'}(x,x') P^{\alpha'}(x')$. 
The bilocal operator $\mathcal{A}$ parallel transports an object at 
point $x'$ to an object at a reference point $x$. One possibility is to define the operator 
as the product of a basis of vector fields (the tetrad fields) at two different points $x$ and $x'$
\begin{equation}
\label{example}
\mathcal{A}^{\alpha}_{\alpha'}(x,x')=e^{\alpha}_{a}(x)e^{a}_{\alpha'}(x'),
\end{equation}
where the latin index $a$ labels the basis vectors.
Then the average of the vector $P^\alpha$ is defined as
\begin{equation}
\langle P^{\alpha}\rangle=\frac{1}{V_{\Sigma}} \int_{\Sigma}d^{4}x'\sqrt{-g'}
\tilde{P}^{\alpha}(x',x),
\end{equation}
where $V_\Sigma$ is the volume of the region $\Sigma$, 
\begin{equation}
V_{\Sigma}=\int_{\Sigma}d^{4}x'\sqrt{-g'}.
\end{equation} 
For higher rank tensors the bilocal extension works on each space-time index and the 
average is defined analogously. Especially this allows us to define the bilocal extension of the metric 
tensor and the definition of an averaged metric.

Zalaletdinov defines a line element for the macroscopic space-time
\begin{equation}
ds^2=\bar{g}_{\mu\nu}dx^\mu dx^\nu,
\end{equation}
where $\bar{g}_{\mu\nu}$ is the macroscopic metric tensor, which is used to calculate the 
macroscopic Christoffel symbols, denoted by $\langle \Gamma^\mu_{\nu\alpha}\rangle$. 
The so defined macroscopic Christoffel symbols guarantee that the macroscopic space-time is a 
Riemannian manifold itself and that there are no metric correlations
\cite{Zalal 2008}. Thus the averaged metric tensor
$\langle g_{\mu\nu} \rangle$ and the averaged inverse metric tensor $\langle g^{\mu\nu} \rangle$ can be 
identified as
\begin{equation}
\langle g_{\mu\nu} \rangle =\bar{g}_{\mu\nu}, \quad \langle g^{\mu\nu}\rangle=\bar{g}^{\mu\nu}.
\end{equation}
 
Now Zalaletdinov defines a macroscopic Riemann tensor
\begin{equation}
M^\mu_{\nu\alpha\beta}=\partial_\alpha \langle \Gamma^\mu_{\nu\beta}\rangle-
\partial_\beta \langle \Gamma^\mu_{\nu\alpha}\rangle +
\langle \Gamma^\mu_{\sigma\alpha}\rangle \langle \Gamma^\sigma_{\nu\beta}\rangle
-\langle \Gamma^\mu_{\sigma\beta}\rangle \langle \Gamma^\sigma_{\nu\alpha}\rangle,
\end{equation}
which is different from the average of the microscopic Riemann tensor 
$\langle R^\mu_{\nu\alpha\beta}\rangle$. 
This non-perturbative approach is also called \textit{macroscopic gravity}. One considers a 
macroscopic description of gravity based on a continuous matter model, instead of a microscopic 
description in which matter would be described by a discrete model.

Following this structure the obtained macroscopic field equations are
\begin{equation}
\langle g^{\beta\rho}\rangle M_{\rho\beta}-\frac{1}{2}\delta^\rho_\gamma \langle 
g^{\mu\nu} \rangle M_{\mu\nu}=8\pi G \langle T^\rho_\gamma \rangle+\langle 
g^{\mu\nu}\rangle (Z^\rho_{\mu\nu\gamma}-\frac{1}{2}\delta^\rho_\gamma Z^\alpha_
{\mu\nu\alpha}),
\end{equation}
where
\begin{equation}
Z^\alpha_{\beta[\gamma \underline{\nu}\sigma]}=\langle \Gamma^\mu_{\beta[\gamma}
\Gamma^\mu_{\underline{\nu}\sigma]}\rangle-\langle \Gamma^\alpha_{\beta[\gamma}
\rangle \langle \Gamma^\mu_{\underline {\nu} \sigma]}\rangle.
\end{equation}
underlined indices are not included in anti-symmetrization and $Z^\alpha_{\mu\nu\beta}
=2 Z^{\alpha \quad \rho}_{\mu[\rho \quad \underline{\nu}\beta]}$. 
Some solutions of macroscopic field equations have been studied in 
\cite{2005, Coley, hoogen}.

A technically simpler approach has been proposed by Buchert. He decomposed Einstein equations 
into a set of dynamical equations for scalar quantities. The disadvantage of this approach is that 
the set of scalar equations is not closed and an assumption like an effective equation of state 
has to be introduced. Despite of its limitation for using only scalars,
Buchert's formalism is the only formalism apart from Zalaletdinov's macroscopic gravity,
that treats the inhomogeneities in an exact way, and gives new insight.

The metric can be written in the synchronous gauge
\begin{equation}
ds^{2}=-dt^{2}+ ^{(3)}g_{\mu\nu}dx^{\mu}dx^{\nu},
\end{equation}
where $^{(3)}g_{\mu\nu}$ is the metric on hypersurface of constant $t$.
The spatial average of a scalar quantity $f$ is defined on these hypersurfaces as
\begin{equation}
\langle f \rangle (t, \textbf{x}) = \frac{\int d^{3}x \sqrt{g(t,\textbf{x})}
f(t,\textbf{x})}{\int d^{3}x 
\sqrt{g(t,\textbf{x})}}.
\end{equation}
 
The averaged scale factor is defined via the comoving volume on spatial hypersurfaces 
\begin{equation}
a_{\mathcal{D}}(t)=(\frac{\int d^{3}x \sqrt{g(t,\textbf{x})}}{\int d^{3}x \sqrt{g(t_{0},
\textbf{x})}})^{\frac{1}{3}}.
\end{equation}

A key point here is that the time evolution and spatial averaging do not commute
\begin{equation}
\partial_t \langle f \rangle -\langle \partial_t f\rangle=\langle f \theta \rangle -
\langle f\rangle \langle \theta \rangle,
\end{equation}
where $\theta=(\sqrt{^{(3)}g})^{-1} \partial_t(\sqrt{^{(3)}g})$ is the expansion rate. 
By considering the expansion shear tensor $\sigma_{\mu\nu}$,
a kinematic backreaction term is defined as $\mathcal{Q}_{\mathcal{D}} \equiv \frac{2}{3}(\langle \theta^2
\rangle -\langle \theta \rangle^2)-2\langle \sigma^2\rangle$. For 
a vanishing cosmological constant 
and irrotational dust (with mass density $\rho$) this leads to Buchert's 
equations (see e.g. \cite{Review})
\begin{equation}
3\frac{\dot a_{\mathcal{D}}^2}{a_{\mathcal{D}}^2}=8\pi G \langle \rho \rangle_{\mathcal{D}}-\frac{1}{2}\langle \mathcal{R}\rangle_{\mathcal{D}}-
\frac{1}{2}\mathcal{Q}_{\mathcal{D}},
\end{equation}
\begin{equation}
3\frac{\ddot a_{\mathcal{D}}}{a_{\mathcal{D}}}=-4\pi G \langle \rho\rangle_{\mathcal{D}}+\mathcal{Q}_{\mathcal{D}},
\end{equation}
where $\mathcal{R}_{\mathcal{D}}$ denotes the 3-Ricci scalar and 
$\mathcal{Q}_{\mathcal{D}}$ and $\langle \mathcal{R}\rangle_{\mathcal{D}}$ have to obey to
\begin{equation}
a_{\mathcal{D}}^{-2}\big(a_{\mathcal{D}}^{2} \langle \mathcal{R}\rangle_{\mathcal{D}}\dot{\big)}+
a_{\mathcal{D}}^{-6}\big(a_{\mathcal{D}}^{6}\mathcal{Q}_{\mathcal{D}}\dot{\big)}=0.
\end{equation}
The most important result of Buchert's approach is that it is possible to cast a spatially
volume averaged irrotational dust model in the form of a FL model with an effective mass
density and pressure.
 
We notice that Buchert's formalism has been considered in
FL space-time where the Weyl tensor vanishes and thus
bundles of light rays are subject to Ricci focussing
(i.e.~associated with a smooth distribution of matter).
However, in a clumpy universe, light rays propagate in
underdense regions and are sensitive to Weyl focussing
(i.e.~induced by the gradient of the gravitational potential).
The issue of relating Weyl focussing of point like sources
to Ricci focussing of smooth matter sources has been considered recently in
\cite{1109.2484}.

Recently, Skarke \cite{skarke} realized that one could avoid the non-commutativity of
averaging and time evolution of scalars by utilizing a mass weighted averaging scheme.
Other averaging schemes that invoke the past light cone will be discussed in Sec. 6 of this work.

A problematic aspect of spatial averaging is that we do not observe spatial volumes, but rather 
null volumes and that Buchert's and Zalaletdinov's approaches neglect possible
effects on the propagation of photons.
 
\section{Propagation of light in an averaged space-time}

In standard cosmology the background geometry is used for observing the large scale 
of universe.
Speaking of \textit{background} means the homogeneous isotropic flat FL universe 
with neglecting the details of small scales and local inhomogeneities. 
 
Observational cosmology is based on light trajectories and the paths of light are on 
null geodesics. One of the significant effects of inhomogeneity is on the light 
trajectories. Therefore we should see these effects on observations in the lumpy 
universe.

Some aspects of this are very well understood and studied in great depth, e.g.
CMB photons are related to density fluctuations by the Sachs-Wolfe effect. The integrated
Sachs-Wolfe effect \cite{sachs}, which is caused by gravitational redshift, and the 
gravitational lensing \cite{Lensing} play important roles in interpreting the effects of
light propagation in the universe.

The crucial issue is how we justify the smaller scales to the background geometry 
and transform from lumpy universe to the smoothed one.
 
The key point here will be to use an averaged metric that describes the smoothed 
manifold. This allows us to consider the paths of light propagation in the averaged 
space-time. But the metric tensor cannot be averaged easily, and many current 
approaches of averaging cannot be used to construct such an averaged metric. 
The exception is the averaging procedure defined by Zalaletdinov and we view it as a 
proof of existence of such an average. Thus we are going to assume that the average of a 
metric is a metric. 

As a second critical assumption, we are proposing that the averaged space-time 
agrees perfectly with the causal structure of the microscopic space-time. We think that this 
is a plausible assumption. At least this assumption is implicitly made in modern cosmology 
when it is assumed that the light rays in the Universe that is assumed to be isotropic and homogeneous 
on large scales are null in a FL model. The example of the bilocal extension (\ref{example}) has this 
property. Unfortunately, it is not useful for our purpose as it gives $\langle g_{\mu\nu} \rangle = g_{\mu\nu}$. 
Nevertheless, this proves that at least one bilocal extension that satisfies both required properties exists. 

Let $k^\mu$ denote a null vector field. Its geodesic equation reads
\begin{equation}
\label{eq:2}
k^{\mu}_{,\nu}k^{\nu}+\Gamma^{\mu}_{\nu\rho}k^{\nu}k^{\rho}=0, 
\end{equation}
where the Christoffel symbol can be calculated from the metric,
\begin{equation}
\label{eq:3}
\Gamma^{\mu}_{\nu\rho}=
\frac{1}{2}g^{\mu\sigma} (g_{\sigma\rho,\nu} + g_{\sigma\nu,\rho} -  g_{\nu\rho,\sigma}).
\end{equation}
We cancel out the inverse metric in (\ref{eq:2}) by a multiplication with $g_{\lambda\mu}$ 
and using 
\begin{equation} 
(k^\mu g_{\mu\nu} k^{\nu})_{,\lambda}=0.
\end{equation}
We arrive at a more convenient form of the geodesic equation for null vector fields, 
\begin{equation}
\label{eq:geo}
(k^{\mu}_{,\nu}g_{\mu\lambda}+k^{\mu}_{,\lambda}g_{\mu\nu}+k^{\mu}g_{\mu\lambda,\nu})k^{\nu}=0 .
\end{equation}
The big advantage for our purpose is that this form is linear in the metric.

As a next step, we average this equation in the following sense: 
We consider a particular light ray, so $k^\mu$ is not subject to the averaging, but the metric 
and it's derivative are. 
Thus we assume that averaging and contractions with $k^\mu$ commute. 
This assumption might not hold for all possible averaging schemes, but we think that this is a sensible 
assumption to make. Finally, we also assume that derivatives of the wave vector are not 
subject to averages. The reason is again that we take to point of view that we only average over the metric 
of space-time, but consider the same light ray in the averaged and microscopic space-time. Thus 
the coordinate derivatives of the wave vector should not be affected by this averaging procedure. In other 
words, the same coordinate values correspond to the same physical event in the averaged and microscopic 
space-times. Distances, angles and time intervals between physical events are different however.

A consequence of the two assumptions mentioned above is that we preserve the null condition,
\begin{equation}
\langle k^\mu g_{\mu\nu} k^{\nu} \rangle =  k^\mu \langle g_{\mu\nu} \rangle k^\nu = 0. 
\end{equation} 
Note that the derivative of an averaged metric is different from the average of the 
derivative of the metric, i.e. $\langle g_{\mu\lambda,\nu}\rangle\neq\langle g_{\mu\lambda}\rangle_{,\nu}$.

Therefore we arrive at an averaged version of (\ref{eq:geo}),
\begin{equation}
\label{eq:modiT}
\left(k^{\mu}_{,\nu} \langle g_{\mu\lambda} \rangle+k^{\mu}_{,\lambda} \langle g_{\mu\nu} \rangle
+k^{\mu}\langle g_{\mu\lambda} \rangle _{,\nu}\right)k^{\nu}=k^{\mu} T_{\mu\lambda\nu} k^\nu \equiv I_\lambda, 
\end{equation}
where $T_{\mu\lambda\nu} \equiv \langle g_{\mu\lambda} \rangle_{,\nu}-\langle g_{\mu\lambda,\nu} \rangle$. 
The left hand side represents the equation of a null geodesic of an averaged metric and the 
right hand side represents the modification due to averaging. 

Let us now prove two important properties. Firstly, it turns out that the object 
$T^{\rm sym}_{\mu\lambda\nu} = (T_{\mu\lambda\nu} + T_{\nu\lambda\mu})/2$ is a tensor. 
This is a non-trivial and non-obvious statement. Secondly, we can show that 
$I_\lambda k^\lambda =0$.

The tensor property follows from a brute force argument, based on the
well know transformation properties of vectors and tensors (the averaged metric has 
been assumed to be a tensor itself)
\begin{equation}
\label{eq:stransf}
k'^{\mu} (x') = \frac{\partial x'^{\mu}}{\partial x^\alpha}k^\alpha(x),  
\end{equation}
and 
\begin{equation}
\langle  g'_{\mu\nu} \rangle (x')= \frac{\partial x^{\alpha}}{\partial 
x'^{\mu}}\frac{\partial x^\beta}{\partial x'^{\nu}} \langle g_{\alpha\beta} \rangle (x) , 
\end{equation}
and their derivatives
\begin{equation}
\frac{\partial k'^{\mu}}{\partial 
x'^{\nu}}=\frac{\partial x^\beta}{\partial x'^\nu}
\frac{\partial x'^\mu}{\partial x^\alpha}\frac{\partial k^\alpha}{\partial x^\beta}
+\frac{\partial x^\beta}{\partial x'^\nu}\frac{\partial^2 x'^\mu}
{\partial x^\alpha \partial x^\beta}k^\alpha, 
\end{equation}
and  
\begin{eqnarray}
\label{eq:etransf}
\!\!\!\!\!\!\! 
\frac{\partial \langle g'_{\mu\nu} \rangle}{\partial x'^\lambda}=
 \frac{\partial x^\sigma}{\partial x'^\lambda}
 \frac{\partial^2 x^\alpha}{\partial x^\sigma \partial x'^\mu}
 \frac{\partial x^\beta}{\partial x'^\nu}
 \langle g_{\alpha\beta}\rangle+
 \frac{\partial x^\sigma}{\partial x'^\lambda}
 \frac{\partial x^\alpha}{\partial x'^\mu}
 \frac{\partial^2 x^\beta}{\partial x^\sigma \partial x'^\nu}
 \langle g_{\alpha\beta}\rangle+
 \frac{\partial x^\sigma}{\partial x'^\lambda}
 \frac{\partial x^\alpha}{\partial x'^\mu}
 \frac{\partial x^\beta}{\partial x'^\nu}
 \frac{\partial \langle g_{\alpha\beta}\rangle}{\partial x^\sigma}.
 \end{eqnarray}
We then check explicitly that the left hand side is a vector and conclude that 
$k^{\mu}k^{\nu} T_{\mu\lambda\nu}$ transforms as a vector. 
For doing so, one has to use the relation
\begin{equation}
 \frac{\partial^2 x'^\mu}{\partial x^\gamma \partial x^\alpha}
 \frac{\partial x^\beta}{\partial x'^\mu} = 
 -\frac{\partial x'^\mu}{\partial x^\alpha}\frac{\partial^2 x^\beta}{
 \partial x^\gamma \partial x'^\mu},
\end{equation}
besides the null condition. Finally we can argue that as $I_\lambda = k^{\mu} T_{\mu\lambda\nu} k^\nu$
is a vector, the symmetric part, $T^{\rm sym}_{\mu\lambda\nu} = 
(T_{\mu\lambda\nu} + T_{\nu\lambda\mu})/2$, must also be a tensor. 
 
Notice that by construction $T_{\mu\lambda\nu}$ is a symmetric object under the exchange of $\mu$ 
and $\lambda$, $T_{\mu\lambda\nu}=T_{\lambda\mu\nu}$, but not necessarily a tensor. 
Also note that $T^{\rm sym}_{\mu\lambda\nu} \neq T^{\rm sym}_{\lambda\mu\nu}$. 
In fact only $T^{\rm sym}_{\mu\lambda\nu}$ is of relevance to the averaged null geodesic equation.

The second property, $I_\lambda k^\lambda = 0$, follows from 
contracting the left hand side of (\ref{eq:modiT}) with $k^\lambda$ and using the fact that the null 
property of the wave vector is preserved. A straightforward calculation shows that the left 
hand side vanishes identically and thus the second statement holds. 
  
\section{Propagation of light through an averaged Universe}

Let us denote the metric of a FL model by $\bar g_{\mu\nu}$ and the four-velocity 
of a comoving observer (the one that sees the light ray under consideration) by $\bar u^\mu$. 
The observed photon frequency is then given by $\omega \equiv -\bar u_{\mu}k^{\mu}$. 

We assume that $\langle g_{\mu\nu} \rangle = \bar g_{\mu\nu}$, as the cosmological 
principle tells us that we should be able to describe the averaged Universe by an isotropic 
and homogeneous model. We still do not define how this average works in detail, but we 
assume that it exists and argue that observations confirm that such an approach must
be possible.   

Applying the cosmological principle to the tensor $T^{\rm sym}_{\mu\lambda\nu}$, we can write 
down the most general algebraic structure compatible with isotropy and homogeneity. It is 
 \begin{equation}
 T^{\rm sym}_{\mu\lambda\nu} = \frac{f_1}{2} (\bar g_{\mu\lambda} \bar u_{\nu} +\bar g_{\nu\lambda} \bar u_{\mu}) 
 + f_2 \bar u_{\mu} \bar u_{\lambda} \bar u_{\nu}+ f_3 \bar g_{\mu\nu} \bar u_{\lambda}, 
 \end{equation}
as $\bar u^{\mu}$ and $\bar g_{\mu\nu}$ are the only non-trivial tensors of first and second 
rank that can be used to construct a third rank tensor that is symmetric in two of its indices.
$f_1, f_2$ and $f_3$ are three functions of cosmic time $t$ only, which cannot be fixed by 
pure symmetry considerations. However, only the combination 
\begin{equation}
I_\lambda = k^\mu T^{\rm sym}_{\mu\lambda\nu} k^\nu =  f_1 (- \omega) \bar g_{\mu \lambda} k^\mu 
+ f_2 \omega^2 \bar u_\lambda, 
\end{equation} 
enters the averaged light geodesic equation. We further consider the contraction
\begin{equation}
I_\lambda k^\lambda =  - f_2 \omega^3, 
\end{equation}
which must vanish as shown in the previous section and thus $f_2 \equiv 0$.  

Thus the inhomogeneity of the light propagation equation is given by 
\begin{equation} 
I_\lambda = - f_1  \omega \bar g_{\mu \lambda} k^\mu, 
\end{equation}
and all effects of averaging on the light propagation must be encoded in a single function $f_1(t)$.
Without any further knowledge, this generic structure of the inhomogeneity of the null 
geodesic equation allows us to make some non-trivial and generic statements about light 
propagation in an averaged Universe. 

The Hubble rate $H = \dot{a}/a$, 
where $a(t)$ denotes the scale factor of the averaged FL metric and the dot denotes a derivative 
with respect to cosmic time. 
For a comoving observer with $\bar u_{\mu}=(-1,0)$ we find $k^{0}=\omega$ and $k^{i}= \omega e^{i}/a$, 
where $e^i$ is a spatial unit vector, indicating the spatial direction the light ray is pointing at.

Let us first look at the time component of the averaged null geodesic equation.
The left hand side is well known from the equation of null geodesic motion in the FL model, 
\begin{equation}
\label{eq:I0}
(-\omega)(\dot{\omega} + \frac{e^i}{a} \omega_{,i}  +  H \omega)= I_0 = f_1 \omega^2.
\end{equation}
The equation might be more familiar in terms of the affine parameter
\begin{equation}
\frac{{\rm D}\,  \omega}{{\rm d} \lambda}  = \frac{{\rm d}\,  \omega}{{\rm d} \lambda} = k^\mu \frac{\partial\, \omega}{\partial x^\mu} 
= \omega (\dot{\omega} + \frac{e^i}{a} \omega_{,i}), 
\end{equation}
and thus 
\begin{equation}
\label{eq:effective}
\frac{{\rm d}\,  \omega}{{\rm d} \lambda} +  H_{\rm eff}\,  \omega^2 = 0, \quad H_{\rm eff} \equiv H + f_1.
\end{equation}
For $f_1 = 0$ this reduces to the famous result $\omega \propto 1/a$, the redshift of photons (note that 
${\rm d} t = \omega {\rm d} \lambda$). 
Thus we conclude that any $f_1 \neq 0$ leads to a modification of the redshift of photons, so we would 
expect that the actual redshift of a photon in an averaged description of an inhomogeneous universe 
must differ from the redshift that the same photon would have in the corresponding homogeneous 
and isotropic universe. 

The spatial components of the modified light propagation equation becomes
\begin{equation}
\label{eq:Ii}
a \gamma_{ij} e^j \omega \left(\dot{\omega} + \frac{e^k}{a} \omega_{,k}  +  H \omega \right) +  
a \omega^2 \gamma_{ij}\left(\dot{e}^j + e^j_{|k} e^k \right) = I_i = - \omega^2 a \gamma_{ij} e^j f_1,
\end{equation}
where $\bar g_{ij} = a^2 \gamma_{ij}$ and $|$ denotes a covariant derivative with respect to the 
3-metric $\gamma_{ij}$. By means of (\ref{eq:I0}), equation (\ref{eq:Ii}) can be further simplified to yield
\begin{equation}
\dot{e}^j + e^j_{|k} e^k  = 0, 
\end{equation}
or in terms of the affine parameter
\begin{equation} 
\frac{{\rm D} e^i}{{\rm d}\lambda} = 0,
\end{equation} 
i.e.~light rays propagate along straight lines. This result holds for an exact FL models and for 
equation of motion for light in the averaged Universe. 

To sum up, based on the principles of statistical isotropy and homogeneity, 
there is one global effect on the propagation of light, which is a modification of the redshift of a 
photon, which can be described by an effective Hubble expansion rate. 

\section{Discussion}

In order to estimate the function $f_1(t)$ we consider an irrotational model without gravitational waves. 
Then an ansatz for the metric that allows for density perturbations and can easily be compared with the 
zero shear gauge (or longitudinal Newtonian gauge) of linear perturbation theory is
\begin{equation}
 ds^2= - e^{2\phi} {\rm d}t^2 + a^2(t) e^{- 2 \psi} \gamma_{ij} {\rm d}x^i {\rm d}x^j.
\end{equation}
We split the exact metric $g_{\mu\nu} = \bar g_{\mu\nu} + \delta g_{\mu\nu}$, where we do not make 
the assumption that $\delta g_{\mu\nu}$ is small. By construction $\langle \delta g_{\mu\nu}\rangle = 0$.

We now evaluate
\begin{equation}
 T_{000}^{\rm sym} = -\langle \delta g_{00,0} \rangle=2\langle e^{2\phi} \dot \phi \rangle.
\end{equation}
Note that $\bar g_{00} = -1$ and thus its derivative vanishes before and after averaging. 
By comparing this result with our ansatz for $T_{000}^{\rm sym}=f_1$ we have 
\begin{equation}
 f_1=2\langle e^{2\phi} \dot \phi \rangle.
\end{equation}
Alternatively $\delta g_{ij}=a^2 (e^{- 2 \psi} - 1) \gamma_{ij} \psi$ allows us to estimate $f_1$ from 
$T_{ij0}^{\rm sym}$, where now $f_1= 2\langle e^{2 \psi} \dot \psi \rangle$.
Without anisotropic pressure, the off-diagonal components of the Einstein tensor must vanish, 
which implies $\phi = \psi$ (exact!) and thus both estimates are consistent with each other. 

As already stated above, $\langle e^{2 \phi} \rangle \equiv 1$ (by construction).  
However, since averaging and time derivative 
do not commute in general, $f_1$ is in general non-zero. In linear perturbation theory, $\dot \phi =0$ 
in the Einstein-de Sitter model (EdS), but this is not the case for the $\Lambda$CDM model. 
For higher orders in perturbation, both the EdS and the $\Lambda$CDM model have $\dot \phi \neq 0$ 
(these are the integrated Sachs-Wolfe effect \cite{sachs} and the Rees-Sciama effect
\cite{Rees}). Consequently, this implies that 
for the fully nonlinear theory we have $\langle  e^{2 \phi} \dot \phi \rangle \neq 0$ in general.

Let us now estimate qualitatively what are the effects of the effective Hubble expansion rate $H_{\rm eff}$.  
By means of (\ref{eq:effective}), $H_{\rm eff}=H+2\langle e^\phi \dot \phi \rangle$.
In the following, we define the density contrast w.r.t. the averaged matter density $\bar\rho(t)$, i.e.
\begin{equation}
 \delta(\textbf{r},t)\equiv \frac{\rho(\textbf{r},t)-\bar \rho(t)}{\bar \rho (t)}.
\end{equation}
$\rho \ge 0$ implies $\delta \ge -1$. For an over-dense, collapsing region ($\delta > 0$), we 
expect from Newtonian reasoning that $\dot \phi <0$. Similarly, for an underdense, expanding 
region ($\delta<0$), $\dot \phi >0$. However, if the overdense region is virialized, its gravitational potential 
does not change any more and we expect no effect. Thus it is impossible to predict the sign of 
$f_1$ without a detailed investigation. Another important aspect is that most of the volume of the Universe 
is under-dense. An arbitrary light-ray will typically pass 
through a dominantly under-dense universe, and thus we expect that $H_{\rm eff} > H$ at times long after 
the formation of cosmic structure started. On the other hand, observed light is typically emitted in an 
over-dense and observed in an over-dense region. Thus for objects at not too far distances we expect 
that over-densities dominate the trajectory of the light ray. For a quantitative discussion, which is beyond the 
scope of this work, some numerical simulations are necessary.  

We can nevertheless conclude that the one-to-one association of redshift with the scale factor and thus 
with cosmic time that we know from the standard model of cosmology is not possible if the effect from
the averaged description is taken into account.  

Let us finally put our work in the context of a previous result.
In the work of R\"as\"anen \cite{R2009, R2010} the propagation of a bundle of light has been studied.
The redshift $z \equiv (\omega_{\rm s} - \omega_{\rm o})/\omega_{\rm o}$, where the suffixes denote source 
and observer, is found to be
\begin{equation}
 1+z=\exp(\int^{\lambda_{\rm o}}_{\lambda_{\rm s}} {\rm d} \lambda \omega
 [\frac{1}{3}\theta+\sigma_{\mu\nu}e^\mu e^\nu]), 
\end{equation}
where $\sigma_{\mu\nu}$ denotes the shear and $\theta$ the expansion rate and $e^{\mu}$ denotes as above 
the spatial direction of light propagation. This result agrees very well with our result in equation 
(\ref{eq:effective}), which after integration can be written as (using ${\rm d}t = \omega {\rm d}\lambda$)
\begin{equation}
 1+z=\exp[\int^{t_{\rm o}}_{t_{\rm s}}\! \! H_{\rm eff}\,  {\rm d} t ]. 
\end{equation}
R\"as\"anen argued that the shear is negligible for the averaged geometry, and that the only important 
contribution would come from the averaged expansion rate. Therefore the distance redshift relation
is in terms of the averaged expansion rate. In \cite{R2012} it has been discussed that if the metric
remains close to a FL model, the change in redshift respect to its background value is small. 

\section{Conclusion}
 
In this work, we have considered the propagation of light rays in an averaged space-time.
Our central result is a modification of the equation of null geodesic motion, see (\ref{eq:modiT}).
This new equation of motion is a fully covariant vector equation for the wave-vector $k^\mu$.
Rays describing the propagation of light in an averaged space-time are generated by this
wave vector, which is null w.r.t.~the averaged space-time. In order to prove those points we 
assume that the averaged space-time (pseudo-)metric is a tensor and that it respects the causal 
structure of the microscopic space-time. That such averaging procedures exist has been 
shown by Zalaletdinov \cite{Zalal 2008}. As we consider a fixed light ray (source and observer are 
fixed events on the manifold) we think that it is justified not to average the wave vector and its derivative, 
but to just average the metric and its derivatives.  
 
We then apply this light propagation equation (recall, it is not the geodesic equation of the averaged 
space-time) to a cosmological model. We assume that the averaged metric is a flat, 
spatially isotropic and homogeneous (as suggested by the success of the standard model 
of cosmology). We have shown that the relation between photon frequency and affine parameter
is modified. This modification can be expressed as an effective Hubble rate, as shown 
in (\ref{eq:effective}). Our result is in perfect agreement with previous non-perturbative 
investigations \cite{R2012} and with the results 
of the study of toy models, like the Swiss cheese model \cite{Fleury3}. Also perturbative studies are in line 
with our findings \cite{veneziano5}.

So far we restricted our attention to the study of a single light ray. The next logical step is to study the 
equation of geodesic deviation in order to ask if an analogous modification occurs, which would allow 
us to find a modification to the luminosity and angular diameter distances. In this context it will be interesting to 
ask if it is true that a microscopic Weyl focussing leads to an effective Ricci focusing after averaging.
 
We thus have shown that the Hubble rate associated with the averaged space-time metric does 
not necessarily coincide with the effective Hubble rate that should be considered for photon propagation. 
A quantitative study of the order of magnitude of the effect is beyond the scope of this work. The most 
important result of this work is that the averaging effects on light propagation can be absorbed into an 
effective Hubble rate. This might be one of the more fundamental reasons for the great success of the 
Friedmann-Lema\^itre models.  
 
\section*{Acknowledgments}
We thank Thomas Buchert, Domenico Giulini, Seshadri Nadathur and Harald Skarke for valuable discussions and comments. 
We acknowledge support by Deutsche Forschungsgemeinschaft (DFG) within the 
Research Training Group 1620 ‘Models of Gravity’.
 

\end{document}